\documentclass[
preprint,
amsmath,amssymb,
]{revtex4-2}

\usepackage{graphicx}
\usepackage{dcolumn}
\usepackage{bm}
\usepackage{xcolor}

\begin{document}
\preprint{}
\title{Information dynamics of our brains in dynamically driven disordered superconducting loop networks}

\author{$^{1}$Uday S. Goteti}
 \email{ugoteti@ucsd.edu}   
\author{$^{2}$Shane A. Cybart}
\author{$^{1}$Robert C. Dynes}

\affiliation{$^{1}$Department of Physics, \\ University of California San Diego, La Jolla, CA, 92093, USA.}
\affiliation{$^{2}$Department of Electrical and Computer Engineering, University of California, Riverside, CA 92521, USA.}



\keywords{Information dynamics $|$ Brain network model $|$ Disordered systems $|$ Superconductors }

\begin{abstract}
Complex systems of many interacting components exhibit patterns of recurrence and emergent behaviors in their time evolution that can be understood from a new perspective of physics of information dynamics, modeled after one such system, our brains. A generic brain-like network model is derived from a system of disordered superconducting loops with Josephson junction oscillators to demonstrate these behaviors. The loops can trap multiples of fluxons that represent quantized information units in many distinct memory configurations populating a state space. The state can be updated by exciting the junctions to allow the movement of fluxons through the network as the current through them surpasses their thresholds. Numerical simulations performed with a lumped circuit model of a 4-loop network show that information written through excitations is translated into stable states of trapped flux and their time evolution. Experimental implementation on the 4-loop network shows dynamically stable flux flow in each pathway characterized by the junction firing statistics. The network separates information from multiple excitations into state categories with large energy barriers observed in simulations that correspond to different flux (information) flow patterns observed across junctions in experiments. Strong evidence for associative and time-dependent (short-to-long-term) memories distributed across the network is observed, dependent on its intrinsic and geometrical properties as described by the model. Loop network topology abstraction using the model separates the flowing patterns of information from its physical constraints and describes systems of any scale and complexity. The accuracy of flow statistics are limited by the resolution of local external measuring clock(s) revealing the universal nature of information dynamics through the stated two principles.
\end{abstract}


\maketitle


\section*{Significance statement}
Systems with complex many-body interactions from quantum to large scale are known to exhibit behaviors that cannot be easily computed from fundamental entities. A new perspective of statistical physics relating them to information dynamics of our brains reveals patterns in behaviors independent of their spatiotemporal scales. This paper introduces a mathematically incomplete brain-like network model based on a system of disordered superconducting loops with Josephson junctions. The model embeds within it, the limits of detailed observations into such systems that affect the accuracy of their predictable behaviors. Numerical simulations and experiments demonstrate the model and show properties such as associative and time-dependent memory. It leads to a new approach to design networks inspired by our brains.

\section*{Introduction}
Emergent behaviors exhibited by complex many-body systems characterized by criticality, phase transition etc., in between recurrent patterns are described as the effects of collective spontaneous activity of a large number of its individual constituents and time-dependent interactions between them. They are observed on the system as a whole and cannot easily be derived from a detailed knowledge of the constituents. Such behaviors are the basis for many of their higher-level functions such as memory and cognitive functions of our brains. A new perspective of physics of information dynamics is introduced to understand them starting with an abstraction of these systems into $network$s. A network of coupled non-linear oscillators (e.g., neurons of our brain) representing a disordered system serves as a basis for a generic model. The network evolves through a finite space of all of its observable states as it translates the external excitations to a spatio-temporal distribution of information across it. Long-range interactions between oscillators allow it to span the memory across a wide range of timescales. The excitations dynamically drive the system while it equilibrates with its environment and converges toward a nearest local free-energy minimum that represents a memory state. Closed-loop feedback mechanisms actively reconfigure the energy landscape of the state space to define time-dependent memory. A physical manifestation of a network of elements with short- to long-range interactions can be operated to approximate the dynamic information flow patterns in the network topology of our brains that relate the excitations and the generated responses \cite{goteti2021superconducting},\cite{goteti2021low}. 

Superconducting loops with Josephson junctions provide a nearly ideal model due to macroscopic coherence and quantization of state-space. A minimalist network is demonstrated both experimentally and in simulations. Each superconducting loop can trap multiples of magnetic flux quanta ($\Phi_0$ = $2.067\times10^{15}$ $T/m^2$) or fluxons in the form of either clockwise or counterclockwise circulating supercurrents around the loop. A multi-loop disordered system, schematically shown in Fig. \ref{Fig1}A, can host a large number of distinct flux configurations that are either statically or dynamically stable that populate a discrete state space. A subset of these can be configured to be local energy minima or attractors. Each state defines an interaction strength between pairs of different junctions and any state transition triggers a response that can be measured by the voltage across them. Active external locally applied currents or flux stream excitations drive the system through different states and into the nearest local minimum after they are turned off. A small-scale network of 4 loops was simulated using a lumped circuit model with junctions and inductors to reproduce experimental observations obtained from an equivalent YBCO-based loop network. The following section discusses, through simulations, the operation of a superconducting disordered system and demonstrates addressing the memory states and re-configuring the state space. Then the network model is introduced to relate the simulation and experimental results discussed later. The evolution of the state in time obtained from simulations statistically relates external excitations to locally measured responses across junctions observed in experiments. Time evolution with multiple excitations leads to memory states defining dominant flux flow (information flow) pathways through it such as in our brains. These flux pathways exhibit wide-ranging temporal stability with dynamically changing external excitations depending on the loop geometry and coupling from the relatively stable trapped flux from memory.

\section*{Flux configurations}
The number of all possible trapped flux configurations in a superconducting disordered system is determined by its geometrical or junction parameters and limited by flux quantization. A closed superconducting loop can sustain persistent circulating currents that aggregate to an integer number of fluxons in the absence of any external excitations. Junctions are the weak links through which fluxons can enter or exit a loop generating a voltage response in time (i.e., \(\int_{0}^{t} V \,dt\)=$n\Phi_0$) when an excitation current together with circulating current from previously trapped flux exceeds their critical current barriers. The junctions in a disordered loop system can hence be treated as equivalent to the active neuron-like components of the network interacting through the trapped flux overlapping them. Non-hysteretic (damped) junctions are considered in simulations and experiments that can show single fluxons traversing them in the form of discrete voltage spikes. An integer number of fluxons can be trapped as circulating currents around individual loops (e.g., $n_1\Phi_0$ and $n_2\Phi_0$ around loops $1$ and $2$ in Fig. \ref{Fig1}A) that constrain the interactions to a short range, i.e., between the junctions in that loop. Due to macroscopic coherence across the superconducting system, the network allows long-range direct interactions between junctions through trapped flux where circulating currents can encompass several loops (e.g., shown in Fig. \ref{Fig1}A as $n_{345}\Phi_0$ involving loops $3$, $4$, and $5$). Flux configurations are typically composed of a combination of interactions ranging from short-to-long length scales in a network where interactions (both direct and indirect) between any pair of junctions can be described as due to a cumulative effect of the flux trapped in each closed-loop pathway encompassing either of them. In the rest of the article, the flux configuration or the memory state of the network is defined collectively by the integer number of trapped fluxons $n_k$ for each of the circulating current paths $k$ (e.g., $k=1, 2, 45, 367$, etc.). $n_k$ is considered to be positive for clockwise circulating currents. These configurations can be represented as co-ordinates in an appropriate multi-dimensional energy state space $C$ as shown in Fig. \ref{Fig1}B. Interaction between any pair of junctions can be controlled by traversing the space to a desired state through external excitations that define the dimensionality of $C$.

Fluxon dynamics through the loop network can be modeled using lumped circuit analysis. Its motion through superconducting loops with junctions is comparable to angular strain propagation through torsion bars with suspended pendulums. Pendulums are analogous to junctions and are treated using their circuit equivalents described by the resistively and capacitively shunted junction model. The superconducting paths between the junctions similar to the torsion bar connecting the pendulums are treated using their equivalent inductances as shown in Fig. \ref{Fig1}A, so the angular strain in the torsion bar can be compared to the current through the inductor. The dynamic disordered system therefore behaves as a multi-dimensional array of coupled oscillators interacting over a range of lengths and time scales.

Each circulating current path $k$ can individually accommodate a maximum trapped flux with an integer value $N_k=\frac{L_{k}I_{k}^{j}}{\Phi_0}$, where $L_k$ is the total inductance of the circulating current path and $I_{k}^j$ is the critical current of the junction $j$ where flux traverses out of that path, which is dependent on its interactions with all the junctions in the network. Effects of state fluctuations from thermal noise and quantum tunneling below junction critical currents are neglected in the simulation model. The simulations are still consistent with experiments for statistical analysis of fluxon flow with long-time averaging as described by the network model. The energy coordinate of a configuration as a function of time is calculated from the circuit equivalent as the sum of inductive and potential energies stored in all the circuit components in the network due to cumulative currents across each element from the trapped flux and external excitations. Excitations can be considered as flux flow (voltage spike trains) into the network through junctions or as continuous current signals as shown in Fig. \ref{Fig1}C. The excitations generate a response that can be recorded as quantized flux through junctions with active currents acting as feedback signals (Fig. \ref{Fig1}A). Feedback loops allow us to reconfigure the mapping between input signals and the resulting memory configurations. When driven with continuously changing excitations, the evolution of the system through the dynamic configuration space is observable through the time-dependent voltage signals across different junctions.

\subsection*{Flux configuration mapping in a 4-loop network}
\begin{figure*}[h!]
\centering
\includegraphics[width=1\linewidth]{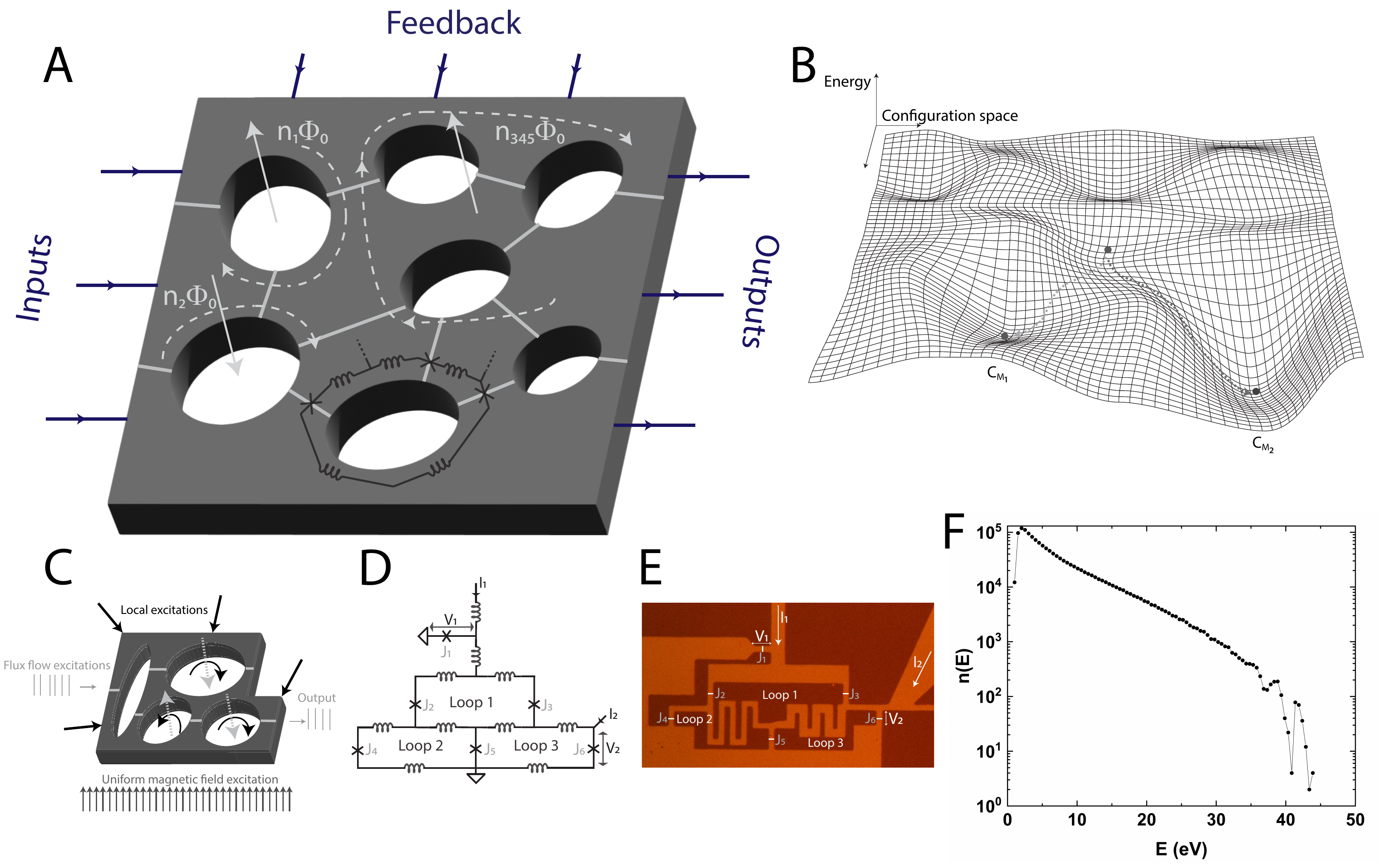}
\caption{\label{Fig1}(A) Superconducting loop disordered network with multiple spiking inputs and outputs. A typical trapped flux configuration with circulating currents around individual loops is shown as $n_1\Phi_0$ and $n_2\Phi_0$ and around multiple loops shown as $n_{345}\Phi_0$. Each loop can be treated in simulations using an equivalent lumped element circuit model of inductors and junctions as shown. Feedback currents are useful to reconfigure the mapping between input signals and the resulting memory states. (B) Configuration space with disordered energy landscape with coordinates defined by all the possible trapped flux states. Stable memory configurations are given by the local energy minima. Any of the trapped flux states can be dynamically reconfigured to be local minima with suitable excitations (inputs or feedbacks). (C) Model of a 4-loop network showing possible ways (arrows) to excite the system to access various flux configurations. (D) An equivalent lumped circuit model of the 4-loop network was implemented to generate the simulation results discussed in the article. (E) YBCO-based 4-loop network with junctions defined by focused He-ion beam irradiation-induced insulating tunnel barriers labeled by white lines. (F) The number of flux configurations and the energy calculated from circulating currents from the circuit model show the density of states from zero trapped flux to maximum filled flux state for the 3-loop memory network.}
\end{figure*}

A simple network of 4 loops with 6 junctions is studied using the circuit model shown in Fig. \ref{Fig1}D and the experimentally equivalent YBCO-based network with junctions as shown in Fig. \ref{Fig1}E. Loops 1 to 3 are constructed to independently accommodate memory states in the form of multiples of fluxons, i.e., $N_{k(=1,2,3)}>1$. Loop 4 is constructed as flux input loop such that $N_{k(=4)}<1$ and any fluxon introduced through junction $J_1$ is stabilized across the 3 memory loops. Disorder is introduced by randomizing the network parameters such as the loop inductances, junction critical currents, and the network topology. The input current $I_1$ induces a voltage $V_{1}$ across junction $J_1$ and the output flux stream is measured as voltage $V_{2}$ across $J_6$ representing a single channel of the network. A second input current $I_2$ is used as feedback to reconfigure the state space relative to the input.

Considering the 3 memory loops, a flux configuration is represented collectively by the respective circulating currents in $7$ different closed-loop pathways as $C = [n_1 \; n_2 \; n_3 \; n_{12} \; n_{13} \; n_{23} \; n_{123}]$. The number of such states as a function of their energy is shown in Fig. \ref{Fig1}F including all possible configurations of trapped flux, calculated by iterating through combinations of $n_k$ up to their maximum $N_k$ that can coexist without surpassing critical currents of any junctions. These configurations comprise most of the available memory states of this 4-loop network. In case of symmetry or order in the individual loops or the coupling between them, iterating through all the possibilities of $k$ results in counting the same state multiple times (e.g., when the same set of circulating currents can be counted twice in $n_{12}$ and $n_{1} + n_{2}$). If the network is completely disordered, the number of possible states increases exponentially with the increasing number of loops in the network and theoretically any of these states can be configured to be meta-stable local energy minima with a complex combination of time-dependent input and feedback signals, as demonstrated in the sections below. 

Experimentally, a transition from any given memory state $C_{M_1}$ to another state $C_{M_2}$ can be initiated by applying multiple different combinations of time-dependent external excitation signals to traverse the dynamic configuration space followed by a relaxation period. The flux configuration at any instant of time can be known with the knowledge of the initial memory state and the flux flow through all the junctions entering or leaving different loops during state evolution. In all the simulations discussed in the article, various states of the 4-loop network are addressed using different excitation signals starting from an initial configuration of zero flux trapped in the system.

\subsubsection*{Magnetic field excitation}
\begin{figure*}[h!]
\centering
\includegraphics[width=1\linewidth]{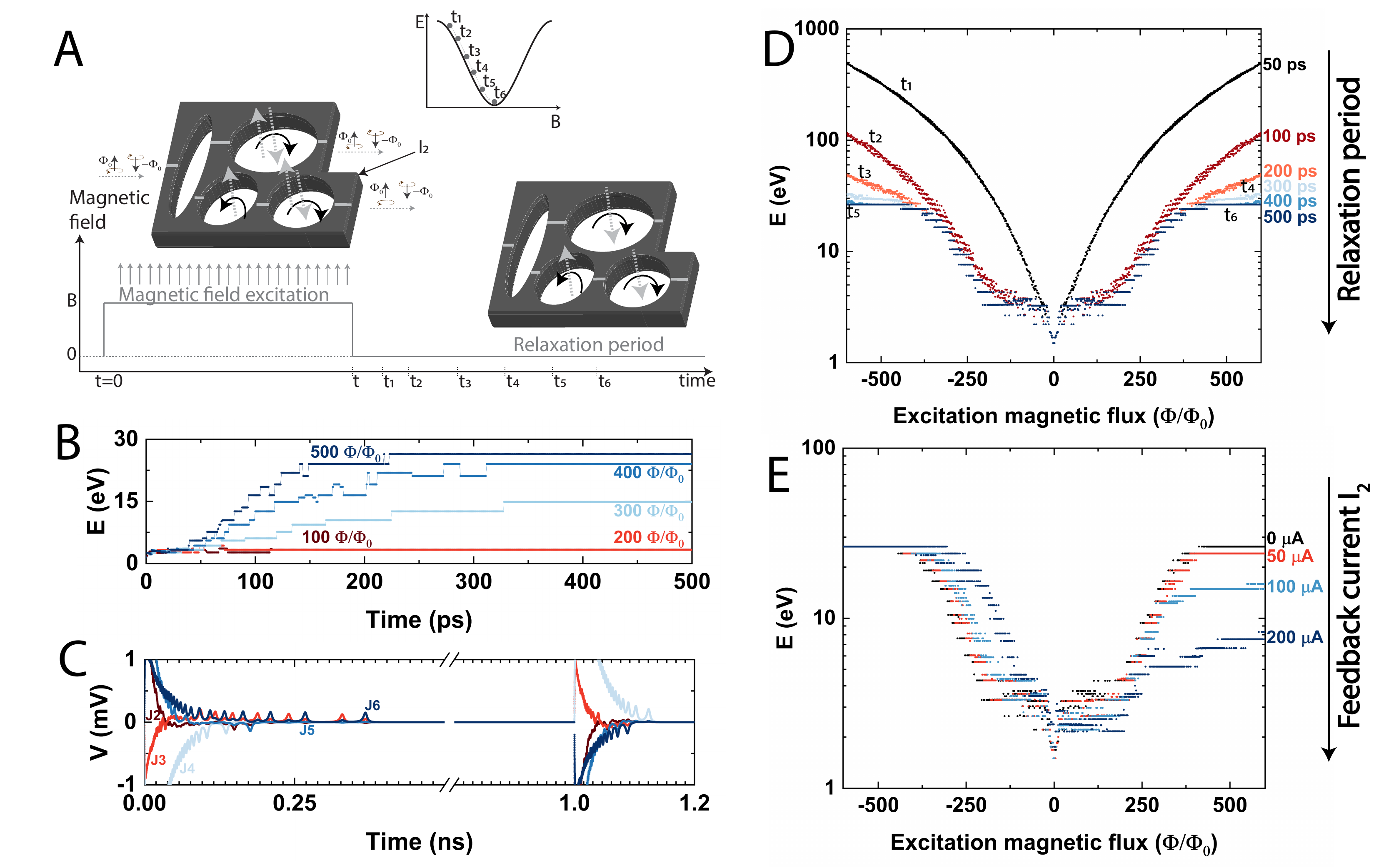}
\caption{\label{Fig2}(A) Schematic illustrating simulations of the uniform magnetic field across the 4-loop network described in Fig. \ref{Fig1}(C-E). A magnetic field pulse is applied with a pulse width of $1ns$ and different pulse heights and then the system is allowed to relax to realize different meta-stable trapped flux configurations. (B) Simulation of time-evolution of the state of the system due to magnetic field excitation. The relaxed state is recorded with pico-second increments to the magnetic field pulse width to map the time evolution for $5$ different pulse amplitudes. This state is independent of time after a critical excitation time period (pulse width) dependent on the loop parameters of the involved trapped flux configurations. (C) The voltage across $5$ junctions in the 3-loop memory network of Fig. \ref{Fig1}(C-E) showing multiples of quantized flux pulses entering and leaving the network through each of the junctions for the magnetic field of $500\frac{\Phi}{\Phi_0}$ with a pulse width of $1 ns$. There is no flux motion after the critical excitation time period. (D) The energy of the state of the network in Fig. \ref{Fig1}(C-E) with the respective magnetic field for a fixed pulse width of $1 ns$ at different times during the relaxation process. Discrete states reached by $1.5 ns$ ($500 ps$ of relaxation) represent the local minima in configuration space. (E) The energy of the relaxed state is shown to demonstrate reconfiguration of the state-space with different constant feedback currents at $I_2$. Feedback current has the effect of tilting the state space to access different memory states.}
\end{figure*}

The flux states of the disordered network can be uniquely addressed by applying a uniform magnetic field across the entire network and allowing it to reach equilibrium before the excitation is turned off. The system will then find the nearest stable local energy minimum. The simulation model shown in Fig. \ref{Fig1}D is subjected to a time-dependent uniform magnetic field through a pulsed current with a short rise/fall time of $1 ps$ across an inductively coupled coil to the outer inductors of the 4-loop network as illustrated in Fig. \ref{Fig2}A. At the onset of the magnetic field pulse, flux begins to enter the network through junctions as the state of the system evolves in time and approaches equilibrium with the external field. If the pulse width of the field is longer than the critical time period required to reach this equilibrium state, the final configuration reached after relaxation is independent of further increase in pulse width as shown in Fig. \ref{Fig2}B. This critical width represents an important timescale during which the effects of dynamics of evolution of the flux state and the excitations cannot be completely separated in the statistical analysis of the observed flux flow responses. Its value is specific to the input excitations and is related to the initial state, the applied field amplitude, and the loop parameters as described using the network model. For the simulated network, the critical excitation time periods for different configurations are found to be in the range of $0-500 ps$ in simulations for different applied fields starting with a zero-flux state. Therefore a pulse width of $1 ns$ is used to map different field strengths to flux configurations as shown in Fig. \ref{Fig2}C-E. Fig. \ref{Fig2}C shows the flux movement in both directions through different junctions in the network during excitation (i.e., from $0-400 ps$) with an equivalent field of $500 \frac{\Phi}{\Phi_0}$ followed by relaxation (from $1 ns-1.2 ns$). The relaxation time is similarly related to the loop parameters. Voltage spikes equivalent to individual fluxons can be observed traversing some junctions (e.g., $J_2$, $J_6$) while a stable flux configuration is reached. 

The energy of the state at different times during the relaxation period following the excitation pulse is shown for different magnetic fields in Fig. \ref{Fig2}D. The state space is symmetric with respect to the sign of the applied field, although the local magnetic moments are expected to be of different directions aligned to the fields. Discrete energy levels of relaxed flux configurations can be seen beginning after $100 ps$, and the network is completely relaxed at $500 ps$ for the entire range of field amplitudes. These states are retained indefinitely in the absence of further positive or negative field excitation. Each state can be addressed uniquely with field amplitudes extending over a range, e.g., the state at $\approx 8 eV$ can be accessed with multiple different field amplitudes in the range of $230 \frac{\Phi}{\Phi_0}$ to $280 \frac{\Phi}{\Phi_0}$. Other states at $\approx 6 eV$, $9 eV$, and $11 eV$ partially overlap with this range of fields, but this overlap is not exact, i.e., any given applied excitation maps to a unique state. These states are adjacent to each other resulting in some uncertainty in the final state in the presence of noise in the excitation. The system reaches a maximum filled flux state along this coordinate axis at a field amplitude of $\approx 450 \frac{\Phi}{\Phi_0}$ and is unchanged with further increase in the field. 

The mapping between the fields and the resulting memory states can be reconfigured through a feedback current applied to the system (shown as $I_2$ in Fig. \ref{Fig1}D, E, and Fig. \ref{Fig2}A). Application of a constant feedback current has the effect of tilting the state-space mapping between the inputs and the final memory states, shown in Fig. \ref{Fig2}E for $I_2$ = $50\mu$A, $100\mu$A and $200\mu$A. States that are not stable previously can be addressed after reconfiguration and are retained even when the feedback current is reduced to zero if the system is allowed to reach equilibrium. In a network with multiple excitations as in Fig. \ref{Fig1}C, some of the available excitation channels can be used as feedback to give access to a large number of memory states shown in Fig. \ref{Fig1}F. A uniform excitation across the entire system in equilibrium conditions therefore shows a simplified time-independent operation of the disordered network where continuous input information can be mapped onto unique discrete states.

\subsubsection*{Local spiking signal excitation}

\begin{figure*}[h!]
\centering
\includegraphics[width=1\linewidth]{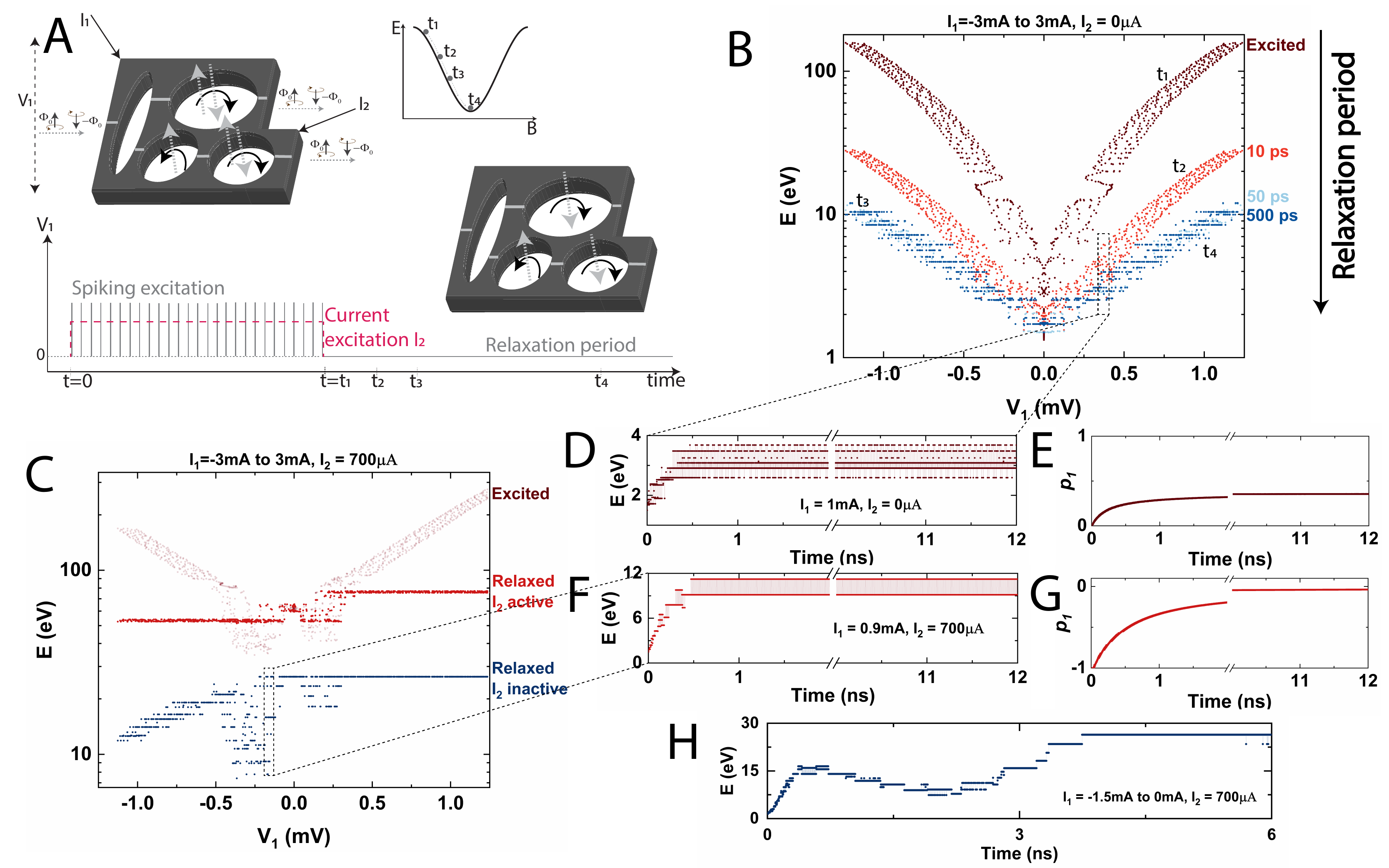}
\caption{\label{Fig3}{\scriptsize(A) Schematic illustrating simulations of spiking excitation induced at $J_1$ in the network described in Fig. \ref{Fig1}(C-E). An input flux flow at of a constant rate is applied for a fixed pulse duration of $t_1=1ns$ for different pulse heights $I_1$ and then the system is allowed to relax to realize different meta-stable trapped flux configurations. (B) The energy of the state of the network in Fig. \ref{Fig1}(C-E) with the respective excitation voltage $V_1$ (i.e., average flow rate of $\frac{V_1}{\Phi_0}$) for a fixed pulse width of $t_1=1ns$ observed at different time intervals (i.e., at $t_2=1.01ns$, $t_3=1.05ns$ and $t_4=1.5ns$) during the relaxation process. Discrete states reached by $1.5ns$ represent the local minima in configuration space. (C) The energy of the relaxed state is shown to demonstrate reconfiguration of the state-space with active constant feedback $I_2 = 700\mu A$ showing network behaviors such as categorization and associative memory. (D) Simulation of time-evolution of the state of the system due to local excitations (i.e., $I_1 = 1mA$ and $I_2 = 0$). The state is recorded after relaxation as a function of pulse width $t_1$ with picosecond increments to map the time evolution. (E) Firing probability $p_1$ of junction $J_6$ with respect to $J_1$ as a function of integration time $T$ during the time evolution of the state shown in Fig. \ref{Fig3}D. (F) Simulation of time-evolution of the state of the system due to local excitations (i.e., $I_1 = 0.9mA$ and $I_2 = 0.7mA$). The state is recorded after relaxation as a function of pulse width $t_1$ with picosecond increments to map the time evolution. (G) Firing probability $p_1$ of junction $J_6$ relative to $J_1$ shown as a function of integration time $T$ during the time evolution of the state shown in Fig. \ref{Fig3}F. (H) Simulation of time-evolution (i.e., the relaxed state after picosecond increments) of the state of the system due to dynamically varying input excitation, i.e., $I_1$ is linearly varied from $-1.5mA$ to $0mA$ in an interval of $6ns$ with constant feedback current $I_2 = 0.7mA$. With time-dependent input, the trapped flux configurations that generate the output flow have different temporal stability.}}
\end{figure*}

Uniform excitations across superconducting system drive it to saturation to address unique time-independent states. Loop networks also allow us to generate local excitations to disordered systems that represent inputs to some neurons to generate local responses to be measured at other output neurons in a network. The 4-loop system of Fig. \ref{Fig1}D and E shows a simplified version with the junctions at the outer-most loop, i.e., $J_1$, $J_3$, and $J_6$ as either inputs or outputs for local excitations or responses. Currents $I_1$ and $I_2$ generate flux flow excitations locally that can be used for either input or feedback. An external current pulse excitation $I_1$ is applied in simulations to induce an input flux flow at a constant rate into the network locally through the junction $J_1$ for a fixed time period as described by Fig. \ref{Fig3}A. Flux flow rate can be quantified from the number of discrete fluxons traversing the junction over a fixed time period, characterized by the constant average frequency or the voltage across it as $\frac{V_{1}}{\Phi_0}$ from the AC Josephson effect. Similar to uniform fields, a sufficiently long period of local flux flow excitation longer than a critical time period allowed the network to reach a dynamic equilibrium. A period of $2 ns$ is observed to be sufficiently long for the network to reach this state giving a steady flux flow into or out of the network through all the junctions on the outer loop. Therefore, a constant flux flow excitation is applied for $2 ns$ and the network is similarly allowed $500 ps$ to completely relax to respective local energy minima for different local flux flow inputs through $V_1$. The flux states shown by their respective potential energy stored are mapped to the input flux flow rates at different instances during the relaxation period using simulations in Fig. \ref{Fig3}B. The state space is qualitatively similar to that of magnetic field excitation but with a larger number of accessible states and a stronger overlap. The final states are dependent on both the flux flow rate and the time period of excitation, as each fluxon during the flow induces a state transition. The configuration map in Fig. \ref{Fig3}B is specific to the chosen $2 ns$. 

Current through $I_2$ can be used as feedback to reconfigure the mapping between input flux flow and the final flux configurations. Simultaneously applying multiple different local excitations (input and feedback) results in network exhibiting brain-like closed loop behaviors. An example considering a constant feedback current of $I_2 = 700\mu A$ is observed to introduce asymmetry to the state space resulting in the separation of states into categories seen as energy contours in Fig. \ref{Fig3}C. Similar behavior is observed for other values of $I_2$ as shown in the experimental results. They can be observed while the excitation currents are on, within regions of $V_1$ between $0$ to $0.3 mV$, $-0.5$ to $-0.25 mV$, etc., and are retained within the same regions even when one or both the currents are turned off. Specifically, the network is in a steady flow state after the input $I_1$ is turned off with $I_2$ turned on and the categories that were originally created with two excitations are retained with energy gaps separating them. A gap of $\approx 15 eV$ separation can be observed at $V_1$ of $-0.1 mV$, and similarly, a gap of $\approx 25 eV$ is seen at $V_1$ of $0.25 mV$. We will find that they represent different flow patterns of flux through the network in experimental observations. Even though the output flow measured after $I_1$ is off is expected to be different, this behavior is characteristic of associative memory between the two inputs $I_1$ and $I_2$. The output flux flow measured before and after $I_1$ is turned off is statistically equivalent (caused by the same category of states), and the flux flow pattern is effectively unchanged as discussed in the sections below.

The discrete states representing uniquely addressable flux configurations in Fig. \ref{Fig2}D, E or the strongly overlapping configurations in Fig. \ref{Fig3}B, C are the quiescent states of the network reached after complete relaxation. Information can only be retrieved through a detailed non-intrusive scan of trapped flux across the entire network. Alternatively, the memory states can be addressed by locally induced flux flow excitations and measuring the resulting flux flow responses, which are a direct consequence of the interactions between the input and the output junctions due to underlying trapped flux memory. This is a destructive readout mechanism as any excitation to the system results in the perturbation of previously written state. A constant flux flow is induced at $J_1$ using current pulses at $I_1$ of increasing excitation pulse widths with pico-second increments, each followed by a relaxation period of $500 ps$. The state evolution in time with each additional input fluxon, obtained from such simulations is shown in Fig. \ref{Fig3}D. With an initial state of zero flux trapped, the state traverses through the configuration space until it reaches a dynamic equilibrium by $400 ps$ as it converges on to a small region observed between $2 eV$ and $4 eV$. The network appears to mimic stochastic fluctuations but exhibits recurrence of states in this region at short time scales. The fluctuations are observed to be deterministic and re-traceable and observed in the absence of noise due to the nonlinearity of the junctions and the disordered coupling between them. In dynamic equilibrium as the network is confined to a subset of recurring states, the interaction strength between the input junction $J_1$ and the output junction $J_6$ is dependent on the stable local minima that pin the flow pattern. An approximate location of the memories in the vicinity of these states can be experimentally measured from the relative flux flow rates, observed over a period to obtain the firing probability of a junction relative to another. Firing probability can be quantified as the number of fluxons exiting $J_6$ due to a fixed number of fluxons entering $J_1$ integrated over a time period $T$ i.e., $p_1 = \frac{\int_0^T{V_2 dt}}{\int_0^T{V_1 dt}}$. $p_1$ is shown as a function of integration times starting from $t=0$ during the time-evolution of the state (Fig. \ref{Fig3}D) in Fig. \ref{Fig3}E. This quantity is independent of the excitations and characterize the flow pattern. The flow probability reaches a constant value after an integration time longer than the critical excitation time period. It is independent of time sufficiently long after the system reaches a dynamic steady state and the network settles into a flow pattern. Another example of dynamic evolution of the state that results in a different firing probability between $J_1$ and $J_6$ with both $I_1$ and $I_2$ on is shown in Fig. \ref{Fig3}F, G. When input flux flow is constant, the dynamic memory state of the system within its vicinity in $C$ is permanently stable. However, a much more general description of a network includes networks subjected to time-varying input and feedback excitations. Such an example is shown in Fig. \ref{Fig3}H with a constant $I_2$ but linearly increasing input current $I_1$ over a period of $6 ns$. The state of the system follows the variations in input as it evolves, but each of the underlying trapped flux states has different temporal stability relative to the changing input excitation resulting in shorter and longer-term dynamic memories. e.g., the trapped flux state between $2.8ns$ and $3.2ns$ is unchanged even as the excitation is varied.

\begin{figure*}[h!]
\centering
\includegraphics[width=1\linewidth]{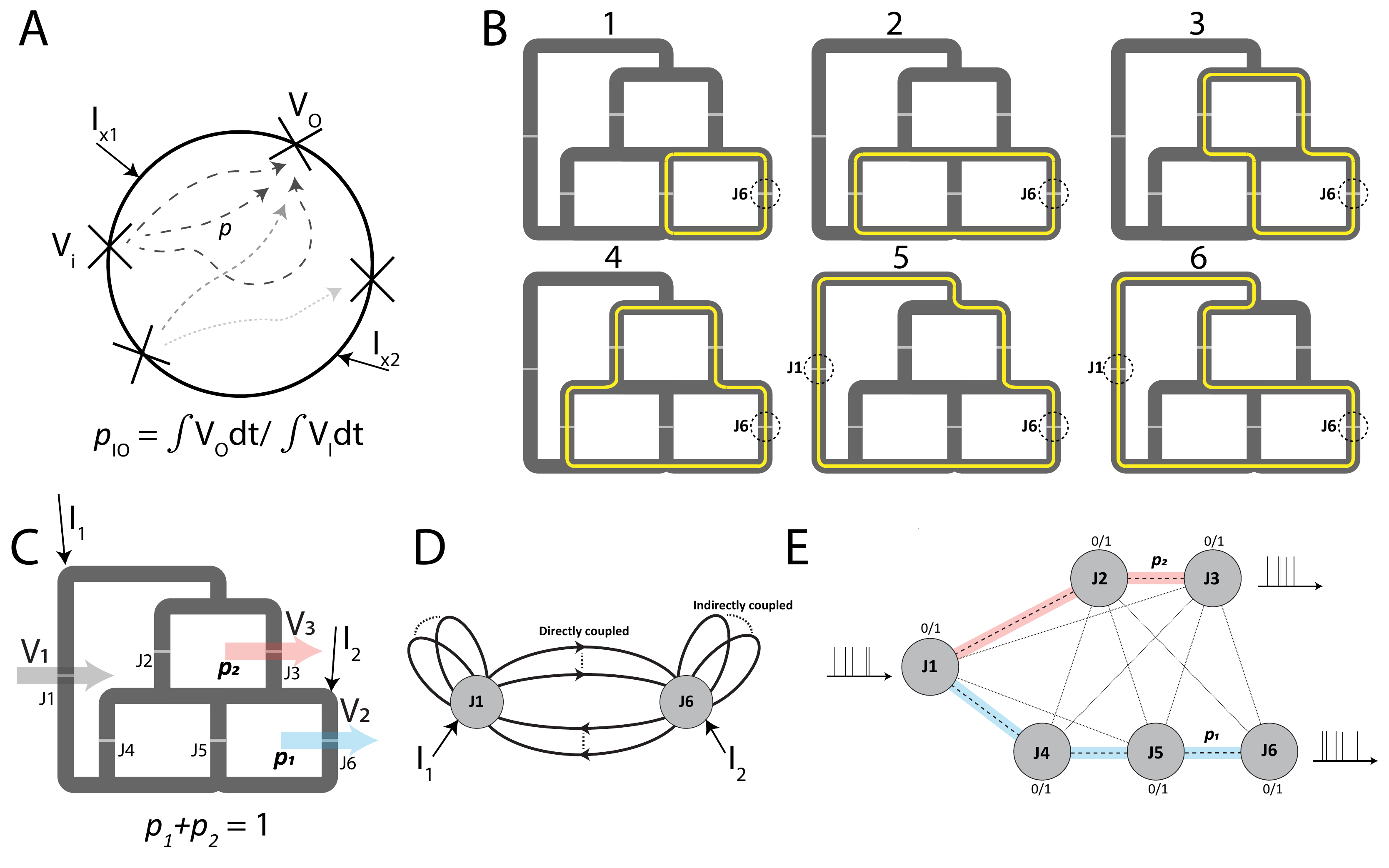}
\caption{\label{Fig4}(A) Graphical illustration of the flux flow patterns between junctions on the outer loop that are defined by the underlying flux configurations that are stable over the integration period. The flux configurations are directly related to the relative junction firing probabilities. (B) A representation of distinct circulating current (trapped flux) paths in the 4-loop network that contribute to the switching activity of the output junction $J_6$. Current paths from 1 to 4 show the contributions from the 3-loop memory while paths 5 and 6 show directly coupled paths between the input junction $J_1$ and the output junction $J_6$. Flux configuration of the network is composed of combinations of flux from all the paths. (C) Input and output flux streams can be measured as voltages $V_1$ (input), $V_2$ and $V_3$ (outputs) across $J_1$, $J_3$ and $J_6$ respectively. Output spiking probabilities are constrained by the sum rule $p_1 + p_2 \approx 1$ for long integration times in a steady state. (D) Correlation between activities of any pair of junctions can be described as due to transitions between directly coupled closed loops including both the junctions, together with transitions in closed loops encompassing either of the junctions. For the 4-loop network, the indirectly coupled loops encompassing only $J_6$ refer to the trapped flux in the 3-loop memory network shown in Fig. \ref{Fig4}B (1-4). All the loops are coupled to each other and $p_1$ is affected by transitions between any of these paths. (E) Network topology representation of the 4 disordered loops with nodes referring to junctions highlighting the dominant flux flow pathways. Two distinct information flow paths ($p_1$ and $p_2$) exist from input to outputs.}
\end{figure*}

\section*{Network model}
The circuit model evaluates a disordered network as an ideal deterministic system so that the time evolution can be accurately predicted when the details of the network parameters and the inputs are all known. It requires a sufficiently high time resolution to track each state transition e.g., at the junction plasma frequency or higher. Simulations of such non-linear dynamic systems of any size, are known to be complex to calculate for longer periods. Experimentally, the evolution can be precisely observed by tracking each discrete transition through measurements of single fluxon movement through all the junctions in the network with a similar time resolution. These observations also become impractical, as the networks become larger and are subject to continuously varying excitations. Alternatively, the approximate state of the network within the vicinity of neighboring states can be experimentally observed statistically from correlations between junction activities obtained by accumulating their firing statistics over a time period $T$, as shown in simulations for steady flow. This analysis is practically feasible and robust to noise as only some of the junctions are required to be monitored. The junctions that can be measured for flux flow with a precision limited by the time resolution (i.e., $T$). Information flow patterns can be mapped on to the network over a wide-range of $T$, which is determined by observational or experimental limits (an external clock). The dependence on $T$ in transition from a deterministic to statistical analysis is the key aspect of information dynamics that relates it to our brains, i.e., the idea of linear synchronous time is acquired from external sources. 

Superconducting loop networks allow a mathematically incomplete representation of spatiotemporal information across the network. Input information is continuously translated from external excitations to flux flow through some (input) junctions that then interact with the previously trapped flux and generate a flow across other (eventually output) junctions in the network. The dynamics of flux entering or leaving the junctions on the outer-most loop of the network (the input or the output junctions) is related to the state evolution in configuration space that can be described by conserving the number of fluxons in the network. This relation is described below by equating the rate of change of total trapped flux in the loop network (i.e., in all the sub-loops $k'$ contained within an outer loop $k$) observed at intervals of integration period $T$, to the flux flow rates measured across the junctions on the outer loop $k$ during this period:

\begin{equation}
\label{Eqn1}
\begin{aligned}
    \sum_{k' \subseteq k}{(n_{k'}(T)-n_{k'}(0))} = \frac{1}{2\pi}{\int_0^T(\sum_{i_k}\dot{\phi_{i_k(t)}})dt}
\end{aligned}
\end{equation}

Flux flow is characterized by the voltage across the junctions, but the phase description is used since the above relation is also true for quantum fluctuations between flux states. The interaction strengths between a pair of junctions on the outer loop at any time can be estimated from their statistical relation as the probability of firing of one (output) junction due to a fluxon induced at another (input), i.e., $p_{IO}=\frac{\int_0^T V_{O}dt}{\int_0^T V_{I}dt}$. When the network is in a dynamic equilibrium, flux configuration is constrained to within its neighboring states and the change in the trapped flux is negligible (the left side of the above equation $\approx0$). This is observed in simulation examples in Fig. \ref{Fig3}D-G and also shown in experimental results discussed later. The firing probability $p_{IO}$ is related to the state (energy) i.e., the flow pattern in the network but independent of $T$. The stable flux configuration in this steady state directs the incoming flux through different pathways to the outputs as it fluctuates locally at shorter timescales than $T$, to generate flux (information) flow patterns of different relative strengths defined in the network topology as shown in Fig. \ref{Fig4}A. The minimum required number of junctions to be monitored to estimate the approximate state from the patterns is therefore equal to the number of junctions on the outer loops when [\ref{Eqn1}] is applied to it and the loops it contains. Detailed patterns can be mapped across the entire network topology, by monitoring the junctions in the other loops where multiple flux flow paths cross each other. 

In a more general case of network operation, input excitations are continuously changing with time, the flux configuration evolves as shown in Fig. \ref{Fig3}H and the network deviates from dynamic equilibrium. The rate of change of trapped flux in the network is finite and sometimes comparable to the flow rate of flux on the outer-loop junctions. The firing probabilities are no longer independent of time or the number of state transitions considered. For any choice of $T$ defined by an external clock, the states that are stable during this period resemble statistics during dynamic equilibrium with steady flux flow patterns. Short-to-long-term memories can therefore be defined relative to this clock. e.g., when the rate of change of trapped flux in loops is very small compared to the flow rate of outer-loop junctions, the states represent a long-term dynamic memory. States that are stable for a shorter duration relative to $T$ (i.e., when the rate of trapped flux change is comparable to the output flow rates) cannot be identified from the junctions on this outer loop alone and require additional junctions with suitable shorter integration periods. The timescales of flux motion between loops that limit the precision of the evolving flow patterns depend on the junctions in the path and their interactions with other junctions or external excitations. 

When a fluxon is induced with local excitation across a junction, it can flow through one of the possible pathways in the network while perturbing previously trapped flux with discrete state transitions. Any state transition is accompanied by a firing event across one or several junctions in that path. Each junction can be described to be in one of the two distinct states in time (i.e., 1: at a finite voltage and $firing$ when $I > I_{k}^j$ and 0: at zero voltage or $idle$ when $I < I_{k}^j$) depending on if the current $I$ through it is either above or below its critical current $I_k^j$ for junction $j$ in loop $k$. Each junction is coupled to all other junctions in the network through one or more closed-loop pathways that can encompass circulating currents from trapped flux. The firing rate of a junction $j$ is proportional to the current $I$ through it given by the cumulative circulating currents from trapped flux $n_k(t)$ in all the loops $k$ through it in addition to the currents from external excitations ($x$) as:

\begin{equation}
\label{Eqn2}
\begin{aligned}
    I_j = (\sum_{k \in j}{\frac{{n_k(t)}{\Phi_0}}{{L_k}}} + {\sum_{x}{c_x^j(t)I_x(t)}})
\end{aligned}
\end{equation}

$c_x^j$ describes a fraction of the excitation $I_x$ through the junction $j$ that depends on the network parameters, the flux configuration $n_k(t)$, and also the excitation currents $I_x(t)$. While the loop network is a discrete state system, it emulates an analog network due to statistical analysis. Even a qualitative description of a junction state as in [\ref{Eqn2}] is useful to estimate the temporal stability of a particular dynamic memory localized to a small category of states. The rate of change of trapped flux in any loop $k$ is given by the derivative of $n_k(t)$. Therefore, when the flow patterns are stable, $n_k(t)$ is unchanged (from equation [\ref{Eqn1}])) and the flow rate through the junctions follows the excitations $I_x(t)$ (from [\ref{Eqn2}]) to produce constant firing probabilities $p_{IO}$. The temporal stability of any single flux flow pathway in the network depends on the relative dynamics of the first and the second terms of [\ref{Eqn2}] to the current $I$ through that junction in time intervals of $T$. For stable excitations, $n_k(t)$ converges to a constant value on the timescale of the loop parameters for charging (or discharging) of flux related to their time constants $\frac{L_k}{R_k}$. $L_k$ is the inductance of loop $k$ and $R_k$ is the equivalent resistance of the junctions in the normal state in that loop. Since $L_k\propto\frac{l_k}{w_k}$, where $l_k$ is the length and $w_k$ is the width of the loop branch, larger loops with longer-range spatial interaction between junctions have higher time constants. They can store longer-term memories relative to smaller loops. This model clearly suggests that the patterns defined in the network topology are independent of their own spatial or temporal scales and are defined only relative to $T$. A complete evaluation of the time-dependent flow patterns can therefore be performed using frequency spectrum analysis of the junction flux responses relative to the excitation spectrum.

Networks can be controlled with excitations and generated responses. Complex operations can be separated into a series of information $write$ and $read$ operations in time. A $read$ operation can be performed with suitable input excitations when the memory state is weakly perturbed within its vicinity during this time period (or the left side of equation [\ref{Eqn1}] $\approx0$) over long integration times. Similarly, a $write$ operation can be performed with input flux flow to induce large changes in the flux configuration relative to feedback currents. Flow at a faster rate relative to the time constants can perform an effective $write$.

The model can be used to map or direct the flux flow patterns in a dynamically evolving network with feedbacks as described in the 4-loop network. The firing activity of the output junction $J6$ (Fig. \ref{Fig4}C) is dependent on the current through it from the trapped flux in loop pathways encompassing it, some examples of which are shown in Fig. \ref{Fig4}B, together with the excitation currents. Loops that are labeled $1$ to $4$ in Fig. \ref{Fig4}B contribute to trapped flux in the 3 memory loops while the examples in Fig. \ref{Fig4}B$5$ and Fig. \ref{Fig4}B$6$ show trapped flux current paths that encompass both the input $J1$ and the output $J6$. Assigning positive sign for flux flow from left to right aligned with clockwise circulating currents, an input flux stream induced at $J1$ can flow towards the other junctions on the outer loop $J3$ and $J6$ that can be observed from relative firing probabilities given by $p_1 = \frac{\int_0^T{V_2 dt}}{\int_0^T{V_1 dt}}$ and $p_2 = \frac{\int_0^T{V_3 dt}}{\int_0^T{V_1 dt}}$ as shown in Fig. \ref{Fig4}C. In a steady flow state with stable flux paths, [\ref{Eqn1}] can be rewritten as $p_1 + p_2 = 1$ and the flow through $J6$ is statistically equivalent to flow through $J4$ and $J5$. Similarly, the flow through $J2$ and $J3$ are equivalent.

Flux flow strength and its temporal stability between two junctions (i.e., $J1$ and $J6$ here) depend on direct interactions between the two junctions through closed-loop pathways in common along with indirect interactions between either of the junctions and the remaining loops or excitations in the network. The input $J1$ and the output $J6$ are directly coupled through common closed loops (e.g., Fig. \ref{Fig4}B loops $5$,$6$) and indirectly coupled through other closed loops (e.g., Fig. \ref{Fig4}B current loops $1-4$ for junction $J6$) that are schematically shown in Fig. \ref{Fig4}D. Each of these pathways represent different length and hence timescales for interactions. The state of the junction $J_6$ and its firing probability $p_1$ with respect to input at $J_1$ is strongly dependent on the indirectly coupled (memory) loops ($k = 1,2,3$) due to their larger size and flux trapping capacity.  Excitation current $I_1$ is strongly coupled to input junction $J_1$ and $I_2$ to output junction $J_6$ due to their short interaction length scales. $I_2$ is treated as feedback to guide the flux flow from input to either of the paths $p_1$ or $p_2$. The probability of a fluxon induced at $J_1$ through the path $p_1$ is higher if the flux is concentrated in loops $2$ and $3$ relative to loop $1$ and aligned to the flow direction. When the excitations are changing at long timescales as in the experiments reported, the temporal stabilities of the flow patterns are dependent only on the loop geometries. For weak excitations i.e., at low currents at $I_1$ and $I_2$ relative to the circulating current from trapped flux, the flux trapped in the 3 memory loops strengthens or weakens the path $p_1$ for output $J_6$. $p_1$ can be programmed by filling the loops (both directly and indirectly coupled) up to a desired memory configuration with large local current excitations over short timescales relative to the time constants of the loops.  Flow patterns are mapped along different pathways with time-dependent strengths $p_1 (t)$ and $p_2(t)$ in the network topology as shown in Fig. \ref{Fig4}E from statistical correlations between junction firing activities using a qualitative analysis of the network topology.

\section*{Experimental results}
\begin{figure*}[h!]
\centering
\includegraphics[width=1\linewidth]{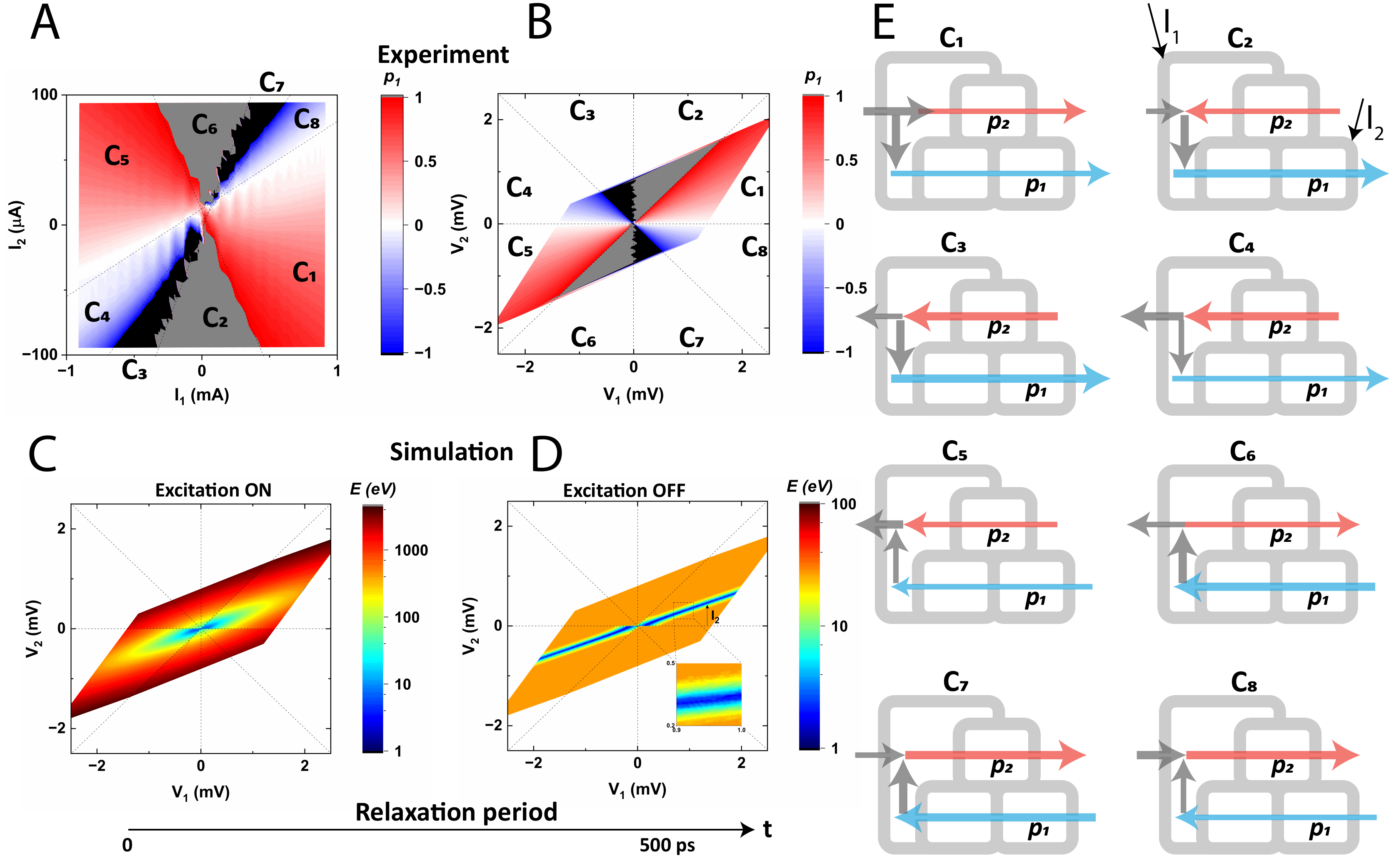}
\caption{\label{Fig5}(A) Flow strength in pathway $p_1$ in the range $-1$ to $1$ as a function of excitation currents $I_1$ and $I_2$. Transitions between flow patterns are disordered reflecting the network structure. Oscillations represent differences in temporal stabilities of states discussed in the text. (B) Flow strength in pathway $p_1$ in the range $-1$ to $1$ across the measurement state-space of $V_1$ and $V_2$ shows $8$ flow patterns and the transition edges between them. (C) Simulation results of the energy of the state with active excitations across the state space of $V_1$ and $V_2$. Energy monotonically increases with $V_1$ and $V_2$ due to the modulation from increasing excitation currents. (D) Simulation results of the relaxed state energy after excitations are turned off, plotted against the state space of $V_1$ and $V_2$ while they were active in a steady state. Saturated states in orange represent the saturated trapped flux in the 3-loop memory also shown in Fig. \ref{Fig6}A. Useful memory states store an output firing probability in the range $0>p_1>1$. (E) Flux flow patterns between input and output are mapped from the experimental results on the 4-loop YBCO network (Fig. \ref{Fig1}E) for a long integration time of $T = 1$ms. Different flow patterns characterize the memory state categories separated either by a change in flow direction along a path or large changes in relative strengths.}
\end{figure*}

The experimental data on the 4-loop network with time-dependent local excitations can be understood from this model. A fabricated YBCO loop network is shown in Fig. \ref{Fig1}E. Measurements were made in a range of temperatures and we show the data collected at 28K. The methods used for fabrication and characterization of the network including early experimental results shown in this paper are reported in \cite{goteti2022superconducting}. A current excitation $I_1$ is used to induce an input flow $V_1$ across $J_1$ and the output flow is measured as $V_2$ across $J_2$, which can be modified using feedback current $I_2$ (Fig. \ref{Fig1}E). The excitations are varied over long timescales with sinusoidally varying $I_1$ and $I_2$ at a few Hz to kHz which are several orders of magnitude slower than the plasma frequencies ranging from $10^9$ to $10^{12}$ Hz that define the flow rate. Therefore, the network is in a dynamic steady state at each data point in all the presented data. The output firing probability is given by $p_1 = \frac{V_2}{V_1}$. To scan the parameter space, the excitation currents $I_1$ and $I_2$ are systematically varied over a range ($I_1$ in $\pm 1 mA$, $I_2$ in $\pm 100 \mu A$). The flux flow probabilities $p_1$ are shown for different input and output flow rates ($V_1$ and $V_2$) in Fig. \ref{Fig5}A with an experimental integration time of $T\approx 1 ms$ per data point. $p_1$ can be varied smoothly at these timescales by varying inputs $I_1$ and $I_2$ but any particular value of $p_1$ can be achieved at multiple different combinations of $I_1$ and $I_2$. This provides a strong experimental evidence that the information flow patterns are independent of the excitation magnitudes or $T$. Additional data in \cite{goteti2022superconducting} consistent with the model show that this is also true for dynamic (sinusoidal) excitations, a characteristic of recurrence in time evolution. When one of the currents is large with respect to the other, $p_1$ is generally same along the line that approximates a constant value of $\frac{I_1}{I_2}$. Exceptions are when the values of $I_1$ and $I_2$ are close such that $T$ is comparable to loop time constants, causing oscillations in $p_1$ along the transition between $C_1$ and $C_8$ (or between $C_4$ and $C_5$) in Fig. \ref{Fig5}A. 

As $I_1$ and $I_2$ are varied to traverse the state space, transitions can occur between states separated by larger energy gaps as shown in the simulations. These transitions are characteristic of large changes to the flux configuration causing dynamically changing flux flow patterns in the network. $8$ regions (labeled $C_1$ to $C_8$) represent different stable flux flow patterns through the network. They can be understood by re-plotting the firing probabilities in the state space of input and output flow rates (i.e., $V_1$ and $V_2$) as shown in Fig. \ref{Fig5}B. The state transitions occur either along $V_1$, $V_2 = 0$ or along $V_1 = V_2$ (i.e., $p_1 = 1$). These transitions are therefore a result of a change in the direction of flux flow (caused by changes in trapped flux) through one or more loops as mapped in Fig. \ref{Fig5}E using Fig. \ref{Fig5}B. Transitions from $C_8$ to $C_1$ or $C_4$ to $C_5$ occur due to a switch in the flux flow direction through $J_1$ or at $V_1 = 0$. Similarly transitions between $C_2$ and $C_3$ or $C_6$ and $C_7$ occur due to flux flow direction change at $J_6$ or at $V_2 = 0$. The flux direction changes between $C_1$-$C_2$ or $C_5$-$C_6$ are at $J_3$ along $V_1=V_2$.

The experiments can be reproduced in simulations to map the energy of the state across the state-space of $V_1$-$V_2$ as shown in Fig. \ref{Fig5}C when excitation is on and in Fig. \ref{Fig5}D when excitation is off to let the network completely relax. State transitions between $8$ different flow patterns are characterized by contours in energy modeling the experiments. The state categories seen in Fig. \ref{Fig4}C overlap with these different flow patterns. The energy of the state is modulated by the excitations in comparison to trapped flux currents due to their relatively larger contributions and is therefore increasing for higher currents (or flux flow rates) observed in Fig. \ref{Fig5}C. When allowed to relax, the underlying stable flux configurations can be realized in Fig. \ref{Fig5}D. It is clear from both experimental and simulation results that when the network is in a steady flow state, the observed firing probabilities in a pathway depend on the flux configuration but are independent of time or the number of firing events, consistent with the network model. Exceptions to these are when the excitation strengths (i.e., flow rates) or timescales match the flow and timescales of flux through loops. States can be programmed at these timescales. Any state can be accessed with multiple different combinations of input flux flows and feedback currents as shown in Fig. \ref{Fig5}D but over different timescales. 

From Fig. \ref{Fig5}D, most of the stable states are confined to a narrow band in regions $C_1$ and $C_5$ for flux input from $J_1$ distributed along the 3 memory loops. Both regions span the region of $p_1$ between $0$ and $1$ but the flow in $C_5$ is inverted in direction. The transition from $C_8$ to $C_1$ is due to the change in the circulating current direction in both loops $2$ and $3$ combined. Similarly, $C_1$ to $C_2$ is due to a flip in the trapped flux direction in loop $1$. The remaining $6$ regions represent a single saturated flux state that is stable over the entire region. The differences in the flow rates in these regions are due to the relative changes between $I_1$ and $I_2$. These saturated states on either extreme bound the range of information in excitations that can be stored in unique states. Therefore they can be accessed with large values of one excitation ($I_1$) with respect to another ($I_2$) or when $\frac{I_1}{I_2}\approx0$ or $\infty$. Along a line of a specific value of $\frac{I_1}{I_2}$ at the transition between $C_1$ and $C_8$, the strength of the pathway $p_1$ is not independent of the scale of excitations but instead shows oscillations. These oscillations span all the flux configurations that cause the flow direction switch through loops $2$ and $3$. At low excitation currents, the contribution to switching from the circulating currents is larger (from [\ref{Eqn2}]). Therefore the oscillation width (equivalent to wavelength) appears to decrease at higher currents. There are significant differences in the interaction length scales between $J_1$-$J_3$ and $J_1$-$J_6$ due to disorder. The oscillations seen relative to a linearly changing excitation are a direct consequence of differences in the temporal stabilities of these different states. Criticality can be defined when such oscillations causing transition are negligible compared to $T$, while emergent behaviors occur along these transitions at shorter timescales with stochastic fluctuations.

\begin{figure*}[h!]
\centering
\includegraphics[width=1\linewidth]{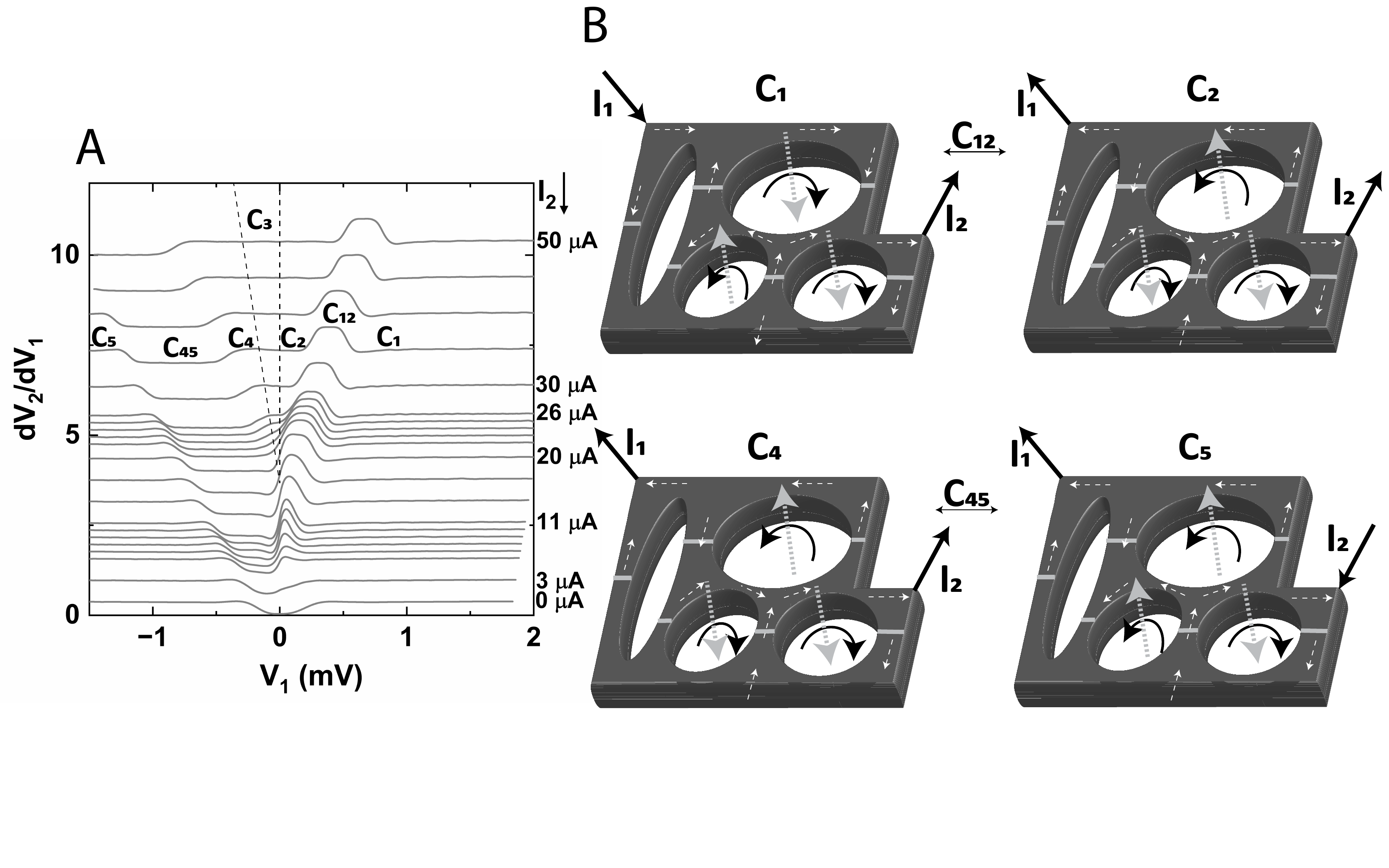}
\caption{\label{Fig6}(A) Measurements of the relative rate of change of flow between $J_1$ and $J_6$ given by $\frac{dV_2}{dV_1}$ for linearly varying input flow $V_1$ with different constant excitations at $I_2$. The transition regions of $C_{12}$ and $C_{45}$ show evidence of time-dependent memory with the transition widths proportional to their temporal stabilities. (B) Trapped flux configurations and the current flow directions for states transitioning from flux flow paths $C_1$ to $C_2$ and $C_4$ to $C_5$. Transition extends over the region of states where the moment of trapped flux in loop 1 is inverted from $C_1$ to $C_2$. Similarly, inversion in flux direction in loops 2 and 3 causes the transition from $C_4$ to $C_5$.  }
\end{figure*}

Dynamically stable (or unstable) states can be experimentally mapped by observing the changes in the relative flow rates observed at the output given by $\frac{dV_2}{dV_1}$. When the network deviates from a steady flow state, the rate of change in trapped flux can be comparable in value to the observed flow rates. At long experimental integration times of $T\approx1ms$, most of the states are unstable and the observed differences in $\frac{dV_2}{dV_1}$ is negligibly small. However, the states that cause changes to trapped flux over long length scales are noticeable due to the long timescales involved in flux movement. $I_1$ is varied between $\pm 1 mA$ for different constant bias currents at $I_2$ and the relative rate of change of flow in the flux path given by $\frac{dp_1}{dt}$ equivalent to $\frac{dV_2}{dV_1}$ is shown for different input flows $V_1$ in Fig. \ref{Fig6}A. Flow patterns from $C_1$ to $C_5$ can be mapped to their respective relative flow rates ($\frac{dV_2}{dV_1}$) for different feedback currents at $I_2$. Even during seemingly abrupt transitions as observed in the state space of $V_1$-$V_2$ (Fig. \ref{Fig5}B), the transition regions are stable over a wide range of $V_1$, labeled as $C_{12}$ between $C_1$ to $C_2$ and as $C_{45}$ between $C_4$ and $C_5$ as shown in Fig. \ref{Fig6}A. The changes to the trapped flux in the 3 loops and the current paths induced by $I_1$ and $I_2$ during these transitions are mapped in Fig. \ref{Fig6}B. Firstly since $C_1$ comprises most of the stable trapped flux states with comparable short-term stabilities, a constant relative rate of flow change $\frac{dV_2}{dV_1}$ is expected. A transition from $C_1$ to $C_2$ involves flipping of trapped flux direction in loop $1$ as shown in Fig. \ref{Fig6}B, during which the flow pattern is stable at $p_1=1$. The region $C_{12}$ spans all the states of loop $1$ from filled to empty while flux in loops $2$ and $3$ is unchanged. When the excitation changes at a much slower rate compared to the timescales of transition through these states, the width of region $C_{12}$ defines the timescale of stability of the flow pattern $p_1=1$ equivalent to the length scale of loop $1$. Similarly, the width of $C_{45}$ is proportional to the length scales of loops $2$ and $3$ combined. The flow pattern of $p_1=0$ ($C_{45}$) is a longer-term memory compared to the flow pattern of $p_1=1$ consistent with the differences in loop inductances. At smaller excitation currents, the changes to the trapped flux are larger relative to the changes to input flow $V_1$. This is also seen in larger oscillations at lower currents in Fig. \ref{Fig5}A. The temporal stabilities of these transition states change relative to excitation currents and can be programmed.
 
\section*{Summary}
This article introduces the idea that any system in nature independent of its scale, complexity, degrees of symmetry or in their absence, as shown in the case of disorder, exhibits universality in patterns of its information dynamics when viewed through an analogy to our brains. The macroscopic state of the system exhibits patterns of recurrence, confining it to a subset of its finite state space that accumulate as memory over time, resulting in emergent behavior out of it. The state evolution establishes statistical relations between external excitations and observed responses, defined in an n-dimensional information space ($I_x$, $x=1,2...n$) over a time sequence, that is mapped onto the system's physical space as its information flow patterns. 

A mathematical model of a generic system follows from network topology abstraction, with nodes across its space acting as gates that measure the flow of information along different pathways. Memory distribution across the system's space is characterized by loops of various sizes that overlap one or several nodes and relate the flowing information to physical constraints of that region. The flow patterns are asynchronous and $time$ can be defined relative to an external local clock measuring the responses. Its $time$ $resolution$ limits the statistical accuracy of the behavior along with the density of nodes throughout the network. The information space of excitations and responses connect the system to a second measuring system of a brain-like network with multiple clocks (coupled oscillators) across it along different timelines. Accuracy of such translation is then limited by the frequency bandwidth and memory capacity of the smaller system. This reveals the universality of information dynamics that relates two or more systems described by the following principles.

\begin{enumerate}
    \item Patterns of information dynamics across systems are independent of its spatio-temporal scales.
    \item Two systems in a closed loop tend towards mimicking each other.
\end{enumerate} 

A complete mathematical description of a system through the above statements requires a continuous distribution of nodes across space requiring expansion of our understanding of topology. They are however demonstrated experimentally with a superconducting loop network with junctions as nodes, as the macroscopic coherence and quantized flux distribution in loops form a minimalistic set of properties for the model. Flux dynamics represents the information flow in the background of the static system of loops. It describes experimental evidence of associative memory at each node where flow patterns cross each other, and time-dependent (short/long term) memory in time sequences determined by the loop size. The model and the underlying theory is also applicable to quantum systems as the dynamics of a fluxon faster compared to time resolution can be explained.  


\section*{Author contributions}
U.S.G and R.C.D. developed the understanding of flux/information dynamics in disordered systems through several blackboard discussions of thought experiments, numerical simulations, and experiments. U.S.G collaborated with S.A.C and R.C.D to perform experiments. U.S.G stated the two principles, developed the theory and the model, performed simulations, and wrote the paper. 

\section*{Acknowledgments}
This work was primarily supported as part of Quantum Materials for Energy Efficient Neuromorphic Computing (Q-MEEN-C), an Energy Frontier Research Center funded by the U.S. Department of Energy, Office of Science, Basic Energy Sciences under Award No. DE-SC0019273.


\end{document}